\newcommand{\be}{\begin{equation}}
\newcommand{\ee}{\end{equation}}
\newcommand{\bea}{\begin{eqnarray}}
\newcommand{\eea}{\end{eqnarray}}

\newcommand{\Appendix}[1]%
    {\renewcommand{\thesection}{Appendix~\Alph{section}:}%
         \section{#1}}%
\catcode`@=11
\long\def\@makecaption#1#2{
   \vskip 10pt
   \setbox\@tempboxa\hbox{{\small\bf #1.} \ {\small #2}}
   \ifdim \wd\@tempboxa >\hsize       
   {\small\bf #1.} \ {\small #2}\par  
   \else                              
        \hbox to\hsize{\hfil\box\@tempboxa\hfil}
   \fi}
\catcode`@=12
\catcode`@=11
\def\secteqno{\@addtoreset{equation}{section}%
\def\theequation{\thesection.\arabic{equation}}}
\def\endsecteqno{\def\theequation{\@ifundefined{chapter}%
{\arabic{equation}}{\thechapter.\arabic{equation}}}}
\newcounter{subequation}
\def\thesubequation{\alph{subequation}}
\def\sneqnarray{\stepcounter{equation}\let\@currentlabel=\theequation
\setcounter{subequation}{1}
\def\@eqnnum{{\rm (\theequation\thesubequation)}}
\global\@eqcnt\z@\tabskip\@centering\let\\=\@eqncr\let\@@eqncr=\@@sneqncr
$$\halign to \displaywidth\bgroup\@eqnsel\hskip\@centering
 $\displaystyle\tabskip\z@{##}$&\global\@eqcnt\@ne
 \hskip 2\arraycolsep \hfil${##}$\hfil
 &\global\@eqcnt\tw@ \hskip 2\arraycolsep
$\displaystyle\tabskip\z@{##}$\hfil
  \tabskip\@centering&\llap{##}\tabskip\z@\cr}
\def\endsneqnarray{\@@sneqncr\egroup $$\global\@ignoretrue}
\def\@@sneqncr{\let\@tempa\relax
   \ifcase\@eqcnt \def\@tempa{& & &}\or \def\@tempa{& &}
   \else \def\@tempa{&}\fi
     \@tempa \if@eqnsw\@eqnnum\stepcounter{subequation}\fi
     \global\@eqnswtrue\global\@eqcnt\z@\cr}
\def\nobiblabels{\def\@lbibitem[##1]##2{\@bibitem{##2}}}
\catcode`@=12
    
\def\e{\eta_b}  \def\u{\Upsilon(1S)}

\documentclass[a4paper,11pt,preprint,aps,prd,showpacs,amsmath,amssymb,nofootinbib,showkeys]{revtex4-1}

\usepackage[german,english]{babel}
\usepackage{hyperref}
\usepackage{amssymb}
\usepackage{amsmath}
\usepackage{bm}
\usepackage{braket}
\usepackage{slashed}
\usepackage{multirow}
\usepackage{epsfig}
\usepackage{epstopdf}

\begin{document}

\preprint{TUM-EFT 70/15}

\title{{\bf Long-range properties of \texorpdfstring{$1S$ bottomonium}{1S bottomonium} states}}

\author{Nora Brambilla}
\email{nora.brambilla@ph.tum.de}
\affiliation{Physik-Department, Technische Universit\"at M\"unchen, \\ James-Franck-Str. 1, 85748 Garching, Germany}
\affiliation{Institute for Advanced Study, Technische Universit\"at M\"unchen, Lichtenbergstrasse 2a, 85748 Garching, Germany}
\author{Gast\~ao Krein}
\email{gkrein@ift.unesp.br }
\affiliation{Instituto de F\'isica Te\'orica, Universidade Estadual Paulista, \\
Rua Dr.~Bento Teobaldo Ferraz, 271 - Bloco II, 01140-070 S\~ao Paulo, SP, Brazil}
\author{Jaume Tarr\'us Castell\`a}
\email{jaume.tarrus@tum.de}
\affiliation{Physik-Department, Technische Universit\"at M\"unchen, \\ James-Franck-Str. 1, 85748 Garching, Germany}
\author{Antonio Vairo}
\email{antonio.vairo@ph.tum.de}
\affiliation{Physik-Department, Technische Universit\"at M\"unchen, \\ James-Franck-Str. 1, 85748 Garching, Germany}

\date{\today}

\begin{abstract}
In the framework of weakly coupled potential nonrelativistic QCD, we derive, first, an analytical expression for the chromopolarizability of $1S$ bottomonium states in agreement with previous determinations. Then we use the QCD trace anomaly to obtain the two-pion production amplitude for the chromopolarizability operator and match the result to a chiral effective field theory with $1S$ bottomonium states and pions as degrees of freedom. In this chiral effective field theory we compute some long-range properties of the $1S$ bottomonium generated by the pion coupling such as the leading chiral logarithm to the $1S$ bottomonium mass and the van der Waals potential between two $1S$ bottomonium states. Both results improve on previously known expressions.
\end{abstract}

\pacs{14.40.Pq, 14.40.Nd, 12.39.Fe, 13.75.Lb}
\keywords{Effective field theories. Bottomonium chromopolarizability. Chiral logarithms. Van der Waals potential.}

\maketitle

\section{Introduction}

The new multiquark XYZ hadrons that have been continuously discovered since the beginning of the last decade~\cite{Brambilla:2014jmp} are the subject of intense study in the literature. They contain a charm quark-antiquark pair, appear near open charm meson thresholds and do not fit with early quark model expectations. Among the various models proposed for the spatial arrangement of the multiquark structure of some of the new hadrons, particularly interesting are those in which the charm quark-antiquark pair remains tightly bound while interacting with the light quarks via multigluon exchanges.  The hadrocharmonium of Ref.~\cite{Dubynskiy:2008mq} is a prominent example of such a model. The multigluon interaction is a QCD analogue of the van der Waals force of atomic physics. In this respect, it is significant that the LHCb Collaboration at CERN reported recently~\cite{Aaij:2015tga} the observation of $J/\psi$-proton resonances in $\Lambda^0_b \rightarrow J/\psi K^{-}p$ decays with properties consistent with pentaquark states of three light quarks and a charm quark-antiquark pair. One conjectured possibility for the structure of the resonances, labeled $P^+_c(4380)$ and $P^+_c(4450)$ by the collaboration, is that of weakly bound molecular states of a baryon and a meson. The latter possibility includes a molecule formed by a light-quark baryon and a charmonium interacting via multigluon exchanges. In fact, some years ago Brodsky, Schmidt and de Teramond pointed out that quarkonium states like $J/\psi$ and $\eta_c$ could form bound states with atomic nuclei due to color van der Waals forces~\cite{Brodsky:1989jd}. A recent lattice QCD calculation by the NPLQCD Collaboration confirms this expectation, finding binding energies of charmonia to light nuclei of the order of a few tens of~MeV~\cite{Beane:2014sda}. By extrapolating their results to physical light-quark masses, the collaboration finds that the binding energy of charmonium to nuclear matter is of the order of $40~{\rm MeV}$ or smaller, in fair agreement with recent model calculations~\cite{deTeramond:1997ny,Krein:2010vp,Tsushima:2011kh,Yokota:2013sfa}. 

A color van der Waals force arises in hadron-hadron interactions due to the chromopolarizability of the color-neutral hadrons, similar to the well-known electric polarizability in atomic physics. Contrary to the situation in QED, not much is presently known about color van der Waals forces; one reason is that they are a long-wavelength feature of QCD and therefore of nonperturbative nature, which makes it difficult to assess them from first principles. The potential relevance of color van der Waals forces for the study of the new hadrons demands a better understanding of their properties within QCD.  Like in many other instances, it is desirable to employ a theoretical framework built on controllable approximations that can be systematically improved. The present paper is a first quantitative attempt in this direction, namely to use the framework of nonrelativistic effective field theories~\cite{Brambilla:2004jw} and chiral effective field theories~\cite{Weinberg:1978kz,Gasser:1983yg} to study long-range properties of the $1S$ bottomonium states. 

$S$-wave quarkonium systems are color neutral and do not possess permanent color-dipole moments or higher-multipole moments. Nevertheless, these states can still interact with gluonic fields through the so-called instantaneous dipole moments. These are created when the quarkonium emits a gluon transitioning into a virtual color-octet state followed by an emission of a second gluon and a return to the original quarkonium state. One often refers to this coupling as the polarizability of the system. The set of the two instantaneous dipoles and the propagation of the intermediate color-octet states forms the quarkonium chromopolarizability.

\begin{figure}[ht]
\centerline{\includegraphics[width=.35\textwidth]{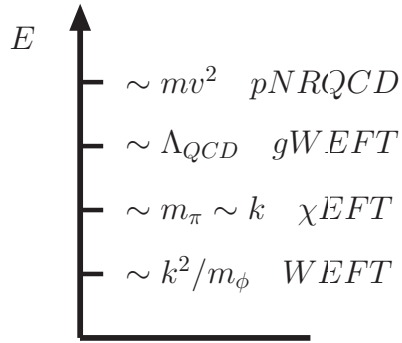}}
\caption{Hierarchy of scales and the corresponding effective field theories;
  $m_{\pi}$ is the pion mass and $k$ is the momentum transfer between two widely separated
  $1S$ bottomonia of mass $m_\phi$ interacting through a van der Waals potential.}
\label{hqcd}
\end{figure}

We will assume that the $1S$ bottomonium states are characterized by the hierarchy of scales $m v\gg m v^2 \gg \Lambda_{\rm QCD}$, where $m$ is the bottom quark mass and $v$ is the relative velocity of the heavy quarks. This assumption stays at the basis of any description of quarkonium as a Coulombic bound state~\cite{Peskin:1979va,Bhanot:1979vb,Voloshin:1978hc,Leutwyler:1980tn}. In the last 20 years, a Coulombic description of the $1S$ bottomonium states has proved to be successful in describing with high accuracy many observables, like electromagnetic, radiative and total widths, hyperfine splittings, etc. It also provides one of the most accurate extractions of the bottom mass. We refer to the reviews~\cite{Brambilla:2004wf,Brambilla:2010cs,Pineda:2011dg} for a compilation of results and an extended list of references.

We can define distinct effective field theories (EFTs) at each scale, see Fig.~\ref{hqcd}. For energies at the soft ($\sim mv$) and ultrasoft ($\sim m v^2$) scales the appropriate EFTs are nonrelativistic QCD (NRQCD)~\cite{Caswell:1985ui,Bodwin:1994jh} and potential nonrelativistic QCD (pNRQCD)~\cite{Pineda:1997bj,Brambilla:1999xf} respectively. One can go one step further and integrate out the ultrasoft scale leading to an EFT with $1S$ bottomonium states and gluons as degrees of freedom at the energy scale of $\Lambda_{\rm QCD}$. We call this EFT, describing color-neutral bottomonium interacting with gluons, gluonic van der Waals EFT (gWEFT). The chromopolarizabilities of the $1S$ bottomonium states are defined in gWEFT and can be computed as matching coefficients. An important element in the calculation of the polarizability is the characterization of the intermediate octet states and their corresponding wave functions. In weakly coupled pNQRCD, the potential of the octet Hamiltonian is a repulsive Coulomb potential, and therefore the octet eigenstates correspond to Coulombic continuum eigenstates.

In the long range, gluons are no longer perturbative and hadronize into pions. Using the QCD trace anomaly we hadronize the two-chromoelectric field polarizability coupling and match the result into a chiral EFT made of $1S$ bottomonium states and pions ($\chi$EFT). The chiral EFT is defined at energies of the order of the pion mass $m_{\pi}$ (see Fig.~\ref{hqcd}). In this way we obtain the values of the low-energy constants of the leading operators coupling pions and $1S$ bottomonium states. These couplings can be used to study the long-range properties of the $1S$ bottomonium states of which we present two: the leading chiral logarithm of the $1S$ bottomonium mass, and the long-range van der Waals potential between two $1S$ bottomonium states. The van der Waals potential is defined in an EFT (WEFT) at the scale set by the kinetic energy of the $1S$ bottomonium (see Fig.~\ref{hqcd}). Matching $\chi$EFT to WEFT, the van der Waals potential is obtained from the two-pion loop that carries the long-range dependence of the two-$1S$-bottomonium interaction. We give an explicit expression for the long-range behaviour of the van der Waals potential.
 
The paper is organized as follows. In Sec.~\ref{pNRQCD} we review briefly pNRQCD. In Sec.~\ref{gWEFT} we introduce gWEFT and perform the matching calculation for the polarizability. In Sec.~\ref{chieft} we write down the $\chi$EFT and, using the QCD trace anomaly, we calculate the low-energy constants associated with the leading-order pion coupling. Using $\chi$EFT we obtain the leading chiral logarithm contribution to the $1S$ bottomonium mass in Sec.~\ref{lechilog}. In Sec.~\ref{vdwp} we obtain the van der Waals potential between two $1S$ bottomonium states and, using a dispersive representation, we find an analytical expression for the long-range potential in coordinate space. Finally, we briefly conclude in Sec.~\ref{seccon}.

\section{\texorpdfstring{\lowercase{p}}{p}NRQCD} 
\label{pNRQCD}

The explicit form of the pNRQCD Lagrangian depends on where the nonperturbative scale $\Lambda_{\rm QCD}$ lies in relation to the soft and ultrasoft scales. The weak-coupling regime of pNRQCD occurs when $mv\gg \Lambda_{\rm QCD}$; in this case, integrating out the degrees of freedom at the energy scale $m v$ can be done in perturbation theory. Furthermore, $v$ can be identified with $\alpha_{\rm s}$. It is convenient to change the coordinates of the fields from the positions of the heavy quark and antiquark to the center-of-mass coordinate $\bm{R}$ and the relative coordinate $\bm{r}$ of the heavy $Q\bar{Q}$ system, and decompose the  $Q\bar{Q}$ field into a color-singlet and a color-octet component. The gauge fields do not depend on $\bm{r}$, since the distance between the heavy quarks is of the order of the soft scale, which has been integrated out. This corresponds to a multipole expansion of the gluon fields. In the present work, we will furthermore assume that $m\alpha_{\rm s}^2 \gg \Lambda_{\rm QCD}$, in which case the physics at the ultrasoft scale is perturbative.

The pNRQCD Lagrangian density in the weakly coupled regime at leading order in $1/m$ and at $\mathcal{O}(r)$ in the multipole expansion is~\cite{Pineda:1997bj,Brambilla:1999xf} 
\be
\begin{split}
\mathcal{L}_{\rm pNRQCD}&=\int d^3r\,\text{Tr}\left[S^{\dagger}\left(i\partial_0-h_s\right)S+O^{\dagger}\left(iD_0-h_o\right)O\right] \\
&+gV_A(r)\text{Tr}\left[O^{\dagger}\bm{r}\cdot\bm{E}S+S^{\dagger}\bm{r}\cdot\bm{E}O\right]+\frac{g}{2}V_B(r)\text{Tr}\left[O^{\dagger}\bm{r}\cdot\bm{E}O+O^{\dagger}O\bm{r}\cdot\bm{E}\right] \\
&+ \mathcal{L}_{\rm light}
\,,
\end{split}
\label{pnrqcdla}
\ee
where $S$ and $O$ are the quark-antiquark singlet and octet fields respectively normalized with respect to color. The Lagrangian $\mathcal{L}_{\rm light}$ is the QCD Lagrangian in the Yang-Mills and light-quark sectors. All the gauge fields in Eq.~\eqref{pnrqcdla} are evaluated in $\bm{R}$ and $t$, in particular $G^{\mu \nu}\equiv G^{\mu \nu}(\bm{R},\,t)$, $\bm{E}^i  \equiv G^{i0}(\bm{R},\,t)$, and $iD_0 O \equiv i\partial_0O - g\left[A_0(\bm{R},\,t),O\right]$. The singlet and octet Hamiltonians read (the relative and center-of-mass kinetic energies are shown up to order $1/m$)
\be
h_s=-\frac{\bm{\nabla_r}^2}{m}-\frac{\bm{\nabla_R}^2}{4m}+V_s(r)\,, \qquad
h_o=-\frac{\bm{\nabla_r}^2}{m}-\frac{\bm{D_R}^2}{4m}+V_o(r)\,.
\ee
At leading order the potentials read ($r=|\bm{r}|$): $V_s(r)= -C_F \alpha_{\rm s}(1/r)/r$, $V_o(r) = \alpha_{\rm s}(1/r)/(2N_cr)$, $V_A(r)=1$, and $V_B(r) =1$, with $N_c=3$, $C_F=(N_c^2-1)/(2N_c)$ and, for further use, $T_F=1/2$.

\section{\texorpdfstring{\lowercase{g}}{g}WEFT} 
\label{gWEFT}

In this section we integrate out the ultrasoft scale $m\alpha_{\rm s}^2$ and match pNRQCD to gWEFT, where gWEFT is an EFT at the energy scale of $\Lambda_{\rm QCD}\ll m\alpha^2_{\rm s}$~\cite{Vairo:2000ia}.  At energies much below $m\alpha_{\rm s}^2$, which is the scale of the binding energy, the different singlet states are frozen and should be considered as independent fields. The degrees of freedom of gWEFT are then the singlet eigenstates and the gluonic fields expressed in terms of the chromoelectric field $\bm{E}$ and the chromomagnetic field $\bm{B}$. In this work, we are only interested in the $1S$ color-singlet eigenstates in the bottomonium sector. These are the spin singlet, $\e$, and the spin triplet, $\u$. Spin-dependent interactions are suppressed by the bottom mass and are beyond the accuracy we are aiming at, therefore these two states can be taken as degenerate and we will represent them both with a $0^{-+}$ field $\phi$. 

In the one-$\phi$ sector, when going from QCD to gWEFT, we integrate out the scales $m$, $m\alpha_{\rm s}$ and $m\alpha_{\rm s}^2$, and thus one is able to organize the gWEFT Lagrangian as a series in the ratios $m\alpha_{\rm s}/m$, $m\alpha_{\rm s}^2/(m\alpha_{\rm s})$ and $\Lambda_{\rm QCD}/(m\alpha_{\rm s}^2)$. The Lagrangian reads (see also Ref.~\cite{Luke:1992tm})
\begin{equation}
\begin{split}
L_{\rm gWEFT}=&\int d^3\bm{R}\left\{\phi^{\dagger}(t,\bm{R})\left[i\partial_0  - E_{\phi} + \frac{\bm{\nabla_{\bm{R}}^2}}{4m}
+\frac{1}{2}\beta g^2 \bm{E}^2_a+\cdots\right]\phi(t,\bm{R})\right\} 
+ \mathcal{L}_{\rm light}\,.
\label{aeft1sl}
\end{split}
\end{equation}
The dots stand for higher-order operators. These can either be relativistic kinetic corrections or other operators coupling $\phi$ to gluons. For this latter kind, the next relevant operator to appear is a coupling to chromomagnetic fields proportional to $\bm{B}^2_a$, which can be shown to be $\alpha^2_{\rm s}$ suppressed with respect to the chromoelectric coupling in \eqref{aeft1sl}. Since chromoelectric and -magnetic fields carry color charge, linear terms in these fields are forbidden in Eq.~\eqref{aeft1sl}.

\begin{figure}[ht]
\centering{\includegraphics[width=0.6\textwidth]{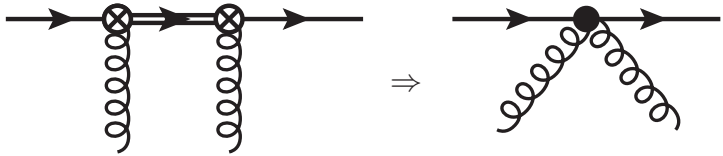}}
\caption{Matching of the pNRQCD diagram on the left-hand side with the gWEFT diagram on the right-hand side. Single lines stand for quark-antiquark singlet and double lines for quark-antiquark octet propagators. The circle with the cross represents the chromoelectric dipole vertex of the pNRQCD Lagrangian \eqref{pnrqcdla}.}
\label{s1m}
\end{figure}

The constant $\beta$ can be termed, in analogy with the electromagnetic properties of neutral systems, as the chromoelectric polarizability. The matching computation for $\beta$ is sketched in~Fig.~\ref{s1m}. The expression for the polarizability reads~\cite{Voloshin:1978hc,Peskin:1979va,Bhanot:1979vb,Leutwyler:1980tn}
\be
\beta=-\frac{2V^2_A T_F}{3N_c} \langle \phi|\bm{r}\frac{1}{E_{\phi}-h_o}\bm{r}|\phi\rangle \,, 
\label{epol}
\ee
where $|\phi\rangle$ is a $1S$ Coulombic state. Note that, since gluons carry color charge, the intermediate states on the left-hand side of Fig.~\ref{s1m} must be color-octet states, this fact is made explicit in the expression of the polarizability in Eq.~\eqref{epol} by the presence of the octet Hamiltonian in the denominator.

\subsection{Polarizability \texorpdfstring{$\beta$}{beta}}
\begin{figure}[ht]
\centering{\includegraphics[width=0.49\textwidth]{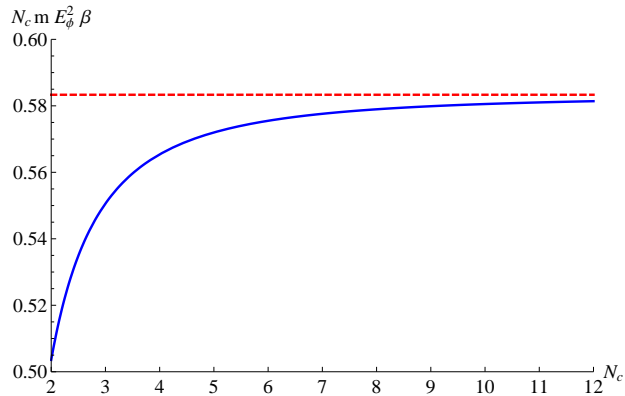}}
\caption{The dependence of the polarizability on the number of colors. The dashed line at the constant value $7/12$ corresponds to the large-$N_c$ limit computed in Ref.~\cite{Peskin:1979va}.}
\label{betaNcplot}
\end{figure}

Octet states can be labeled by their energy and angular momentum quantum numbers and obey
\be
\left(\frac{{\bf p}^2}{m}+V_o \right) |p\,l\,l_z\rangle=\frac{p^2}{m}|p\,l\,l_z\rangle\,.
\ee
It is convenient to introduce an arbitrary unit vector $\hat{\bm{p}}$ and define a state 
\be
|\bm{p}\,l\rangle\equiv \frac{4\pi}{p}\sum_{l_z}|p\,l\,l_z\rangle\langle l\,l_z|\hat{\bm{p}}\rangle \,,
\ee
where $\langle l\,l_z|\hat{\bm{p}}\rangle=Y^{l_z}_l(\hat{\bm{p}})^*$ is a spherical harmonic. A suitable normalization of the states $|\bm{p}\,l\rangle$ is
\be
\sum_l\int\frac{d^3p}{(2\pi)^3}\langle \bm{x}|\bm{p}\,l \rangle \langle\bm{p}\,l |\bm{y} \rangle=\delta^3(\bm{x}-\bm{y})\,.
\ee
By inserting a complete set of states $|\bm{p}\,l\rangle$ into Eq.~\eqref{epol} we get
\be
\beta=-\frac{2V^2_A T_F}{3N_c}\sum_l \int\frac{d^3p}{(2\pi)^3} |\langle \phi|\bm{r}|\bm{p}\,l\rangle|^2\frac{1}{E_{\phi}-\frac{p^2}{m}}\,.
\label{betas2}
\ee
Since $\phi$ is an $S$-wave state, the dipole coupling can only project it into a $l=1$ state due to conservation of the angular momentum. 
Then the only matrix element left to compute is
\be
\langle \phi|\bm{r}|\bm{p}\,1\rangle=\int d^3r\langle \phi|\bm{r}\rangle\bm{r}\langle\bm{r}|\bm{p}\,1\rangle\,.
\ee
The $1S$ (Coulombic) wave function is given by $\langle\bm{r}|\phi\rangle= e^{-r/a_0}/\sqrt{\pi a^3_0}$, with Bohr radius $a_0=2/(m C_F \alpha_{\rm s})$. The Coulombic wave functions in the continuum, $|p\,l\,l_z\rangle$, can be found in Ref.~\cite{abra}, while the octet wave function, $|\bm{p}\,1\rangle$, can be found in Refs.~~\cite{Kniehl:1999ud,Kniehl:2002br,Brambilla:2011sg} and reads
\be
\langle\bm{r}|\bm{p}\,1\rangle=e^{i(\pi/2-\delta_1)}\sqrt{2\pi}\bm{p}\cdot\bm{r}
\sqrt{\frac{\rho\left(1+\frac{\rho^2}{a^2_0|\bm{p}|^2}\right)}{a_0|\bm{p}|\left(e^{\frac{2\pi\rho}{a_0|\bm{p}|}}-1\right)}}
e^{i|\bm{p}||\bm{r}|}\,_1F_1\left(2+i\frac{\rho}{a_0|\bm{p}|};\,4;\,-i2|\bm{p}||\bm{r}|\right)\,,
\ee
where $_1F_1$ is the confluent hypergeometric function, $\delta_1$ is the $l=1$ Coulomb phase and $\rho=(N^2_c-1)^{-1}$. The matrix element squared is then
\be
|\langle \phi|\bm{r}|\bm{p}\,1\rangle|^2=\frac{512\pi^2\rho(\rho+2)^2a^6_0|\bm{p}|\left(1+\frac{\rho^2}{a_0^2|\bm{p}|^2}\right)
e^{\frac{4\rho}{a_0|\bm{p}|}\arctan(a_0|\bm{p}|)}}{\left(e^{\frac{2\pi\rho}{a_0|\bm{p}|}}-1\right)(1+a^2_0|\bm{p}|^2)^6}\,.
\label{mesq}
\ee
Using Eq.~\eqref{mesq} in Eq.~\eqref{betas2}, we arrive at ($V_A=1$ and $E_\phi=-1/(ma_0^2)$)
\be
\beta=256 \frac{\rho(\rho+2)^2}{3N_c}\frac{1}{m E^2_{\phi}}\,I\,,\label{beta2}
\ee
with
\be
I=\int^{\infty}_0dp\,p^3\frac{\left(1+\frac{\rho^2}{p^2}\right)e^{\frac{4\rho}{p}\arctan p}}{\left(e^{\frac{2\pi\rho}{p}}-1\right)
\left(1+p^2\right)^7} ~~\put(11,2){$=$}\put(2,-2){\tiny $N_c=3$}~~~~~~~~ 0.01143\,,
\label{betaI}
\ee
which has been evaluated numerically. The result agrees for $N_c=3$ with Refs.~\cite{Leutwyler:1980tn,Voloshin:1979uv}. Expressions \eqref{beta2} and \eqref{betaI} provide the explicit dependence of the polarizability on the number of colors, see Fig.~\ref{betaNcplot}. For a computation of the polarizability with free wave functions as intermediate states instead of octet ones, which corresponds to the large $N_c$ limit of the matrix element in Eq.~\eqref{epol}, see Refs.~\cite{Peskin:1979va,Bhanot:1979vb}. To our knowledge only this last determination has been used so far in the applications discussed in Secs.~\ref{lechilog} and \ref{vdwp}

\begin{figure}[ht]
\centering{\includegraphics[width=0.49\textwidth]{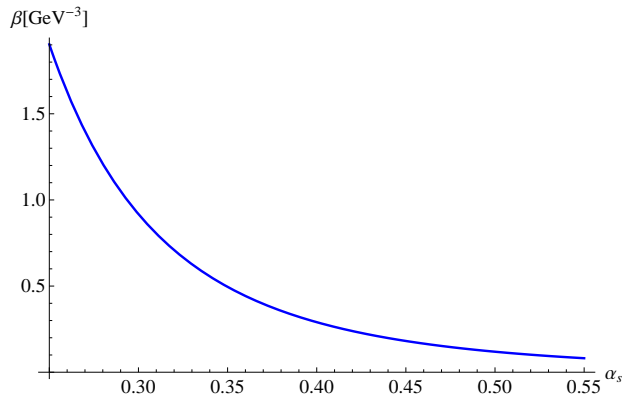}}
\caption{Plot of $\beta$ from Eq.~\eqref{beta2} as a function of $\alpha_{\rm s}$ for $m=5$~GeV.} 
\label{betaplots}
\end{figure}

At leading order, the binding energy, $E_\phi$, is given by $-m(C_F\alpha_{\rm s})^2/4$. In Fig.~\ref{betaplots} we plot $\beta$ from Eq.~\eqref{beta2} as a function of $\alpha_{\rm s}$, with the conventional value of the bottom mass $m=5$~GeV. The natural scale of $\alpha_{\rm s}$ in the binding energy is of the order of the inverse Bohr radius. Taking as the central value for our determination $\alpha_{\rm s}($1.5~GeV$)\approx 0.35$, as the lowest value $\alpha_{\rm s}($2~GeV$)\approx 0.3$ and as the largest value $\alpha_{\rm s}($1~GeV$)\approx 0.5$, we obtain 
\be
\beta=0.50^{+0.42}_{-0.38}~{\rm GeV}^{-3}\,.
\label{betaresult}
\ee
Additional correlated uncertainties come from the bottom mass and higher-order corrections.

We note that by taking the mass of $\phi$ as the spin average of the $\e$~\cite{Tamponi:2015xzb} and $\u$~\cite{Agashe:2014kda} masses, corresponding to the value $m_{\phi}=9.4454$~GeV, and $m=5$~GeV, we get $m_{\phi}-2m=-0.555\,{\rm GeV}$, which is the binding energy we obtain from the leading-order formula when choosing $\alpha_{\rm s} \approx 0.5$. Hence, with this definition of the binding energy, $\beta$ would assume the lowest value in Eq.~\eqref{betaresult}, i.e., 0.12~GeV$^{-3}$. In Ref.~\cite{Voloshin:2004un} the transition $\Upsilon(2S)\rightarrow \Upsilon(1S)\pi\pi$ has been computed using the QCD trace anomaly and the transitional polarizability fitted to experimental data of the decay rates obtaining the value $\beta_{\Upsilon-\Upsilon^{\prime}}=0.66$~GeV$^{-3}$. The same value could be obtained also for $\beta$ using $\alpha_{\rm s} \approx 0.326$. One should notice, however, that $\beta$ and $\beta_{\Upsilon-\Upsilon^{\prime}}$ do not correspond to the same quantity, the latter involving the matrix element between a $1S$ and a $2S$ bottomonium state.

\section{Chiral EFT}
\label{chieft}
At energies of order $m_{\pi}$, much below $\Lambda_{\rm QCD}$, the degrees of freedom are the $\phi$ and the Goldstone bosons. The interaction operators with Goldstone bosons can be easily constructed by considering that the field $\phi$ is a scalar under chiral symmetry. The different sectors of the $\chi$EFT Lagrangian density read at leading order (we include the kinetic energy in Eq.~\eqref{cee})
\bea
{\cal L}_{\rm \chi EFT}^{\phi}&=&\phi^{\dagger}\left(i\partial_0 + \frac{\bm{\nabla^2}}{2m_{\phi}}\right)\phi\,, \label{cee} \\
{\cal L}_{\rm \chi EFT}^{\pi}&=&\frac{F^2}{4}\left\{{\rm Tr}\left[\partial_\mu U \partial^\mu U^\dagger\right]+{\rm Tr}\left[ \chi^\dagger U + \chi U^\dagger\right] \right\}\,,\label{cpipi} \\
{\cal L}_{\rm \chi EFT}^{\phi-\pi}&=&\phi^{\dagger}\phi\frac{F^2}{4}\left\{c_{d0} {\rm Tr}\left[ \partial_0 U \partial_0 U^\dagger \right] +c_{di} {\rm Tr}\left[ \partial_i U \partial^i U^\dagger \right]
+c_{m}  {\rm Tr}\left[ \chi^\dagger U+\chi U^\dagger\right] \right\}\,.\label{cepi}
\eea
The $\phi$ contact interactions are similar to the ones in nuclear physics~\cite{Girlanda:2010ya}. They are essential to renormalize the ultraviolet divergences in the chiral loops, but will not play any role in the long-distance properties that we will discuss in the rest of the paper, hence we do not write them here explicitly. As a basic building block we use the unitary matrix $U(x)$ to parametrize the Goldstone boson fields, which may be taken as
\be
U = e^{i\Phi /F}\,, \quad 
\Phi=\left( \begin{array}{cc}
\pi^0 & \sqrt{2} \pi^+\\
\sqrt{2} \pi^- & -\pi^0
\end{array}
\right)\,,
\label{pions}
\ee
although final results for observable quantities do not depend on this specific choice. At leading order, $F$ may be identified with the pion decay constant $F_\pi=92.419$~MeV. We also use the building block,
\be
\label{blocks}
\chi=2B \hat{m}\bm{1} \,,
\ee
where, working in the isospin limit, $\hat{m}$ is the average quark mass between $m_u$ and $m_d$. The pion mass in the isospin limit is $m_{\pi}^2=2B\hat{m}\approx (135$~MeV$)^2$.

The extension to an $SU(3)$ chiral Lagrangian can be obtained by replacing \eqref{pions} by the appropriate matrix including kaons and etas.

\subsection{Matching gWEFT to \texorpdfstring{$\chi$EFT}{chiEFT} using the QCD trace anomaly}
\label{chiefta}

In the low-energy limit the two-pion production by the polarizability operator of gWEFT in Eq.~\eqref{aeft1sl} is determined up to a constant from chiral algebra and the QCD anomaly in the trace of the energy-momentum tensor~\cite{Chanowitz:1972vd,Chanowitz:1972da,Crewther:1972kn,Freedman:1974gs,Collins:1976yq,Voloshin:1980zf,Novikov:1980fa}. We use this result to match the two-chromoelectric field emission of gWEFT in Eq.~\eqref{aeft1sl} to the pion-$\phi$ interactions in \eqref{cepi}.

\begin{figure}
\centering{\includegraphics[width=0.5\textwidth]{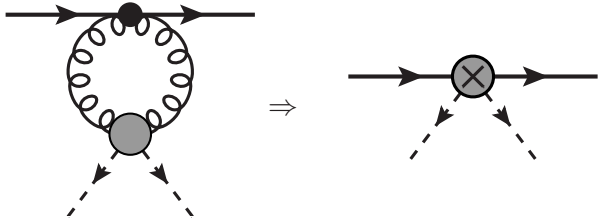}}
\caption{We compute the production of two pions in gWEFT using the trace anomaly (left-hand side) and match the result to the corresponding amplitude in $\chi$EFT (right-hand side). Solid and dashed lines represent $1S$ quarkonium and pions respectively, while the wiggled lines represent gluons.}
\label{s2m}
\end{figure}

Diagrammatically the matching is shown in Fig.~\ref{s2m}. The trace anomaly for the chromoelectric fields is given by \cite{Voloshin:2007dx}  
\be
g^2\langle \pi^+(p_1) \pi^-(p_2)|\bm{E}_a^2|0\rangle=\frac{8\pi^2}{b}\left(\kappa_1 p^0_1p^0_2-\kappa_2 p^i_1p^i_2+3m^2_{\pi}\right)\,,
\ee
where $\kappa_1=2-9\kappa/2$, $\kappa_2=2+3\kappa/2$, $b$ is the first coefficient of the QCD beta function,
\be
b=\frac{11}{3}N_c-\frac{2}{3}N_f\,,
\ee
$N_f$ is the number of light flavors and $\kappa$ is a parameter that can be obtained from pionic transitions of quarkonium states. A detailed experimental study of the decay $\psi^{\prime}\rightarrow J/\psi\pi^+\pi^-$, using the trace anomaly, was done by the BES Collaboration in Ref.~\cite{Bai:1999mj}. The fit to the spectrum of the invariant mass of the produced dipion resulted in the value $\kappa=0.186\pm0.003\pm0.006$, while the fit to the ratio of the $D$- and $S$-wave amplitudes from the angular distribution gave $\kappa=0.210\pm0.027\pm0.042$.

The two-pion production amplitude in gWEFT is
\be
\mathcal{A}=\frac{4\pi^2\beta}{b}\left(\kappa_1 p^0_1p^0_2-\kappa_2 p^i_1p^i_2+3m^2_{\pi}\right),
\ee
which should be matched to the one obtained from $\chi$EFT
\be
\mathcal{A}=-c_{d0}p^0_1p^0_2+c_{di}p^i_1p^i_2-c_m m^2_{\pi}\,, 
\ee
giving
\be
c_{d0} = -\frac{4\pi^2\beta}{b}\kappa_1\,, \qquad
c_{di} = -\frac{4\pi^2\beta}{b}\kappa_2\,, \qquad
c_{m}  =-\frac{12\pi^2\beta}{b}\,. \label{mcm}
\ee

\section{Leading chiral logarithm of the \texorpdfstring{$1S$ bottomonium}{1S bottomonium} mass} 
\label{lechilog}  

\begin{figure}[ht]
\centering{\includegraphics[width=0.5\textwidth]{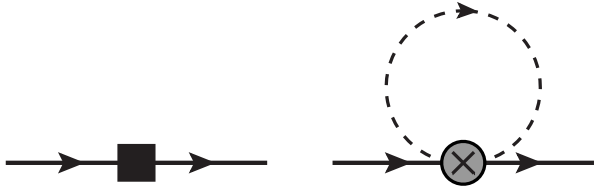}}
\caption{Self-energy contributions to the $1S$ bottomonium mass. The solid square on the left-hand side represents counterterms, while the right-hand side is the pion tadpole diagram that generates the leading chiral logarithm for the $1S$ bottomonium mass.}
\label{selfe}
\end{figure}

One simple and straightforward application of the leading pion-$1S$-bottomonium coupling obtained in Sec.~\ref{chieft} is the determination of the leading chiral logarithm of the $1S$ bottomonium mass. This can be achieved by computing corrections to the $1S$ bottomonium mass up to $\mathcal{O}\left(\beta m_\pi^4\right)$. In Fig.~\ref{selfe} we display diagrammatically the contributions to the $1S$ bottomonium mass. The countertem diagram on the left-hand side contains contributions both at leading order from Eq.~\eqref{cepi} and next-to-leading order from higher-order operators in the chiral Lagrangian that we have not displayed. The tadpole diagram on the right-hand side is constructed only with operators from Eq.~\eqref{cepi}. The mass correction reads
\begin{eqnarray}
\delta m_{\phi}&=& -F^2c_m m_\pi^2  + \hbox{counterterms of}\ \mathcal{O}(m_\pi^4) \nonumber\\
&& + \frac{3 m^2_{\pi}}{8}\left(c_{d0}+3c_{di}-4c_{m}\right)A[m^2_{\pi}]
   + \frac{3m^4_{\pi}\left(c_{d0}-c_{di}\right)}{256\pi^2}\,,
\label{masschirallog}
\end{eqnarray}
where $A$ is the one-point function
\be
A[m^2_{\pi}]=\frac{m^2_{\pi}}{16\pi^2}\left(\lambda-\log\frac{m^2_{\pi}}{\nu^2}\right)\,,
\label{onepoint}
\ee
where $\nu$ is the renormalization scale and
\be
\lambda=\frac{2}{4-d}+1-\gamma_E+\log 4\pi\,;
\label{msbuv}
\ee
$d$ is the space-time dimension. The ultraviolet divergence in $A[m^2_{\pi}]$ can be renormalized in the (modified) $\overline{\hbox{MS}}$ scheme by absorbing the pieces proportional to $\lambda$ in the counterterms.

From Eq.~\eqref{masschirallog}, the chiral logarithm correction to the quarkonium mass reads
\be
\delta m_{\phi}|_{\rm chiral\,log} = - \frac{3}{8}\frac{\beta}{b}m_\pi^4  \log\frac{m^2_{\pi}}{\nu^2}\,,
\label{mchirallog}
\ee
where we have not included chiral logarithms that may be generated from matching $F$ to the pion decay constant beyond leading order. Note that the result does not depend on $\kappa$. A similar approach to ours was used in Ref.~\cite{Grinstein:1996gm} to obtain the light-quark mass dependence of the quarkonium mass splittings. There the polarizability was not computed but was left as a parameter to be fitted on lattice data. Our result disagrees with theirs, which is a factor $16$ larger.\footnote{One source of disagreement can be traced back to a missing factor of $1/2$ when matching the dilatation current with the trace of the chromoelectric field. In contrast to Ref.~\cite{Grinstein:1996gm}, we assert that the contribution from the chromomagnetic field cannot be neglected in that matching.}

\section{\texorpdfstring{$1S$ bottomonium}{1S bottomonium} van der Waals potential} 
\label{vdwp}
In this section, we obtain the van der Waals potential between two $1S$ bottomonium particles. We assume that the momentum transfer $k$ between the two $\phi$'s is of the order of the pion mass, $m_{\pi}$, and therefore the distance between the two $\phi$'s is of the order  $r\sim 1/m_{\pi}$ (the distance $r$ used in this section should not be confused with the quark-antiquark distance defined in Sec.~\ref{pNRQCD}, which is of order $1/(mv)$ and, therefore, much shorter).\footnote{Shorter distance effects would need to be accounted for at the level of pNRQCD, gWEFT or the chiral EFT. They are beyond the scope of the present work.} The van der Waals potential is defined in an EFT (WEFT) at the energy scale of the kinetic operator of the $\phi$ field, which is lower than the pion mass. Hence the potential can be computed as a matching coefficient when integrating out the scale $m_{\pi}$.

\subsection{WEFT}

\begin{figure}[ht]
\centering{\includegraphics[width=0.6\textwidth]{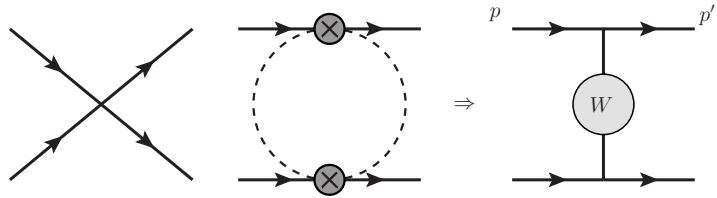}}
\caption{Matching of the amplitude in $\chi$EFT (left-hand side) to a van der Waals potential in WEFT (right-hand side). Solid and dashed lines represent $\phi$ and pions respectively. Neutral as well as charged pions should be considered in the pion loop.}
\label{s3m}
\end{figure}

The energy scale $Q$ of the two-$\phi$ dynamics is given by the kinetic energy of the $\phi$'s in their center-of-mass frame, $Q\sim\mathcal{O}\left(\bm{k}^2/m_{\phi}\right)$, where $\bm{k}$ is the momentum transfer and $m_{\phi}$ is the mass of the $\phi$'s. For a momentum transfer $\bm{k}$ of the order $m_{\pi}$, the interaction of the $\phi$'s is mediated by pions, whose interaction with the $\phi$'s is described by the $\chi$EFT Lagrangian of Sec.~\ref{chieft}. However, since $m_{\pi}\gg m^2_{\pi}/m_{\phi}$ the dynamics of the pions occurs at a higher-energy scale than that of the $\phi$. Therefore, in order to study two-$\phi$ interactions, it is convenient to integrate out the pion dynamics and have its effects taken into account through a potential term. We are going to refer to this term as the van der Waals potential and to the EFT describing the dynamics of $\phi$ interacting through it as WEFT. The Lagrangian of such an EFT is at leading order $L^{\phi}_{\rm WEFT} + L^{\phi\phi}_{\rm WEFT}$, where (we have reabsorbed the mass correction $\delta m_{\phi}$, computed in the previous section, in a field redefinition)
\bea
L^{\phi}_{\rm WEFT}&=& \int d^3 \bm{R} \, \phi^{\dagger}(t,\bm{R}) \left(i\partial_0 + \frac{\bm{\nabla^2}}{2m_{\phi}}\right)\phi(t,\bm{R})\,,
\label{w1e}\\
L^{\phi\phi}_{\rm WEFT}&=& - \frac{1}{2} \int d^3 \bm{R}_1d^3 \, \bm{R}_2\,\phi^{\dagger}\phi(t,\bm{R_1})\,W(\bm{R_1}, \bm{R_2})\,\phi^{\dagger}\phi(t,\bm{R_2})\,.
\label{w2e}
\eea
In Eq.~\eqref{w1e} both the time derivative and the kinetic terms are of the same size $\sim Q$. The potential $W(\bm{R_1}, \bm{R_2})$ can be obtained by matching $\chi$EFT to WEFT, which is shown diagrammatically in Fig.~\ref{s3m}. In the short range it is dominated by contact terms, which include renormalization counterterms. In the long range it depends only on $\bm{k}^2=(\bm{p}-\bm{p}^{\prime})^2$. If we just display the two-pion loop contribution, it reads in momentum space
\begin{eqnarray}
\widetilde{W}(\bm{k}^2) &=& \hbox{contact terms} \nonumber\\
&& -\frac{3}{8} c_{di}\left(c_{d0}-c_{di}\right)\frac{m_\pi^4}{16\pi^2} - \frac{3}{4} c_{di} \left( \bm{k}^2 c_{di} + m^2_{\pi}\left(3c_{di} + c_{d0} - 4 c_m\right) \right) A[m^2_{\pi}] \nonumber\\
&& -\frac{3}{8}\left(\bm{k}^2 c_{di}+2m^2_{\pi}\left(c_{di}-c_m\right)\right)^2 B\left[m^2_{\pi},-\bm{k}^2\right] \nonumber\\
&& -\frac{3}{2}\left(c_{d0}-c_{di}\right)\left(\bm{k}^2c_{di}+2m^2_{\pi}\left(c_{di}-c_m\right)\right) C_1\left[m^2_{\pi},-\bm{k}^2\right] \nonumber\\
&& -\frac{3}{2}\left(c_{d0}-c_{di}\right)^2 C_2\left[m^2_{\pi},-\bm{k}^2\right] \,.
\label{pot}
\end{eqnarray}
$B$ is the standard two-point function, and since $-\bm{k}^2<0$, it takes the form
\be
B\left[m^2_{\pi}\,,-\bm{k}^2\right]=\frac{1}{16\pi^2}\left(\lambda+1-\log\frac{m^2_{\pi}}{\nu^2}+\sqrt{1+\frac{4m^2_{\pi}}{\bm{k}^2}}
\log\left[\frac{\sqrt{1+\frac{4m^2_{\pi}}{\bm{k}^2}}-1}{\sqrt{1+\frac{4m^2_{\pi}}{\bm{k}^2}}+1}\right]\right)\,,
\ee
while the functions $C_1$ and $C_2$ are given by
\bea
C_1\left[m^2_{\pi}\,-\bm{k}^2\right]&=& \frac{5\bm{k}^2+24m^2_{\pi}}{576\pi^2}+\frac{1}{192\pi^2}\Bigg(\left(\bm{k}^2+6m^2_{\pi}\right)\left(\lambda-\log\frac{m^2_{\pi}}{\nu^2}\right) \nonumber \\
&&\left.+\bm{k}^2\left(1+\frac{4m^2_{\pi}}{\bm{k}^2}\right)^{3/2}\log\left[\frac{\sqrt{1+\frac{4m^2_{\pi}}{\bm{k}^2}}-1}{\sqrt{1+\frac{4m^2_{\pi}}{\bm{k}^2}}+1}\right]\right)\,,\\
C_2\left[m^2_{\pi}\,-\bm{k}^2\right]&=&\frac{31\bm{k}^4+280\bm{k}^2m^2_{\pi}+705m^4_{\pi}}{19200\pi^2}+\frac{1}{1280\pi^2}\Bigg(\left(\bm{k}^4+10\bm{k}^2m^2_{\pi}+30m^4_{\pi}\right) \nonumber \\
&&\left.\times\left(\lambda-\log\frac{m^2_{\pi}}{\nu^2}\right)+\bm{k}^4\left(1+\frac{4m^2_{\pi}}{\bm{k}^2}\right)^{5/2}\log\left[\frac{\sqrt{1+\frac{4m^2_{\pi}}{\bm{k}^2}}-1}{\sqrt{1+\frac{4m^2_{\pi}}{\bm{k}^2}}+1}\right]\right)\,,
\eea
where $\nu$ the renormalization scale and $\lambda$ is given in Eq.~\eqref{msbuv}. The ultraviolet divergences can be absorbed in the (modified) $\overline{\hbox{MS}}$ scheme by suitably redefining the counterterms in the first line of Eq.~\eqref{pot} to cancel $\lambda$.

\subsection{Long-range potential in coordinate space} 
\label{lrpot}
In this section, we want to obtain the long-range potential in coordinate space. The potential in coordinate space is the Fourier transform of Eq.~\eqref{pot}. The polynomial terms in Eq.~\eqref{pot} correspond to local Dirac delta potentials and derivatives of it. Since we are interested in the long-range part of the potential we will not consider them. Furthermore, the polynomial part of \eqref{pot} depends on a set of unknown couplings. The long-range part of the potential originates from the pion-loop diagram of Fig.~\ref{s3m}. To obtain the Fourier transform of this piece it is convenient to use a dispersive representation~\cite{Kaiser:1997mw,Epelbaum:2003gr} (we review it in the Appendix). The dispersive representation is useful because it allows us to separate the local from the long-range contributions, namely the subtraction constants give local terms while the two-pion cut gives the long-range contribution. Alternatively one can think of the subtraction constants as redefinitions of the couplings in the polynomial piece of the potential. The potential in coordinate space is obtained through the Fourier transform
\be
W(r)=\int\frac{d^3k}{(2\pi)^3}\,e^{i\bm{k}\cdot\bm{r}}\,\widetilde{W}(\bm{k}^2)\,.
\label{ftp}
\ee
For $\bm{k}^2\rightarrow \infty$ the momentum-space potential $\widetilde{W}(\bm{k}^2)$ diverges as $\bm{k}^4$, and hence its corresponding dispersion relation should be twice-subtracted. Using the spectral representation of Eq.~\eqref{esprep} for $\widetilde{W}(\bm{k}^2)$ in Eq.~\eqref{ftp}, we obtain
\be
W(r)=\frac{1}{2\pi^2 r}\int^{\infty}_{2m_{\pi}}d\mu\,e^{-\mu r}\mu\,\text{Im}\left[\widetilde{W}(\epsilon-i\mu)\right]\,,
\label{coespot}
\ee
where the limit $\epsilon\rightarrow 0$ is understood. The imaginary parts of $B$, $C_1$ and $C_2$ read
\bea
\text{Im}\,B\left[m^2_{\pi},\,\epsilon-i\mu\right]&=&\frac{1}{16\pi}\sqrt{1-\frac{4m^2_{\pi}}{\mu^2}}\,, 
\label{bimp} \\
\text{Im}\,C_1\left[m^2_{\pi},\,\epsilon-i\mu\right]&=&-\frac{1}{192\pi}\mu^2\left(1-\frac{4m^2_{\pi}}{\mu^2}\right)^{3/2}, 
\label{c1imp} \\
\text{Im}\,C_2\left[m^2_{\pi},\,\epsilon-i\mu\right]&=&\frac{1}{1280\pi}\mu^4\left(1-\frac{4m^2_{\pi}}{\mu^2}\right)^{5/2}, 
\label{c2imp}
\eea
from which we obtain the imaginary part of Eq.~\eqref{pot} to be used in Eq.~\eqref{coespot}. We find the following exact expression for the long-range potential in coordinate space:
\bea
W(r)&=&-\frac{3\pi\beta^2m^2_{\pi}}{8b^2 r^5}\left[\left(4\left(\kappa_2+3\right)^2 (m_{\pi}r)^3+\left(3\kappa^2_1+43\kappa^2_2+14\kappa_1\kappa_2\right) m_{\pi} r\right)K_1(2 m_{\pi}r)\right.\notag \\
&&\left.+2\left(2\left(\kappa_2+3\right)\left(\kappa_1+5\kappa_2\right)(m_{\pi}r)^2+\left(3\kappa^2_1+43\kappa^2_2+14\kappa_1\kappa_2\right)\right) K_2(2 m_{\pi} r)\right]\,,
\label{cspot2}
\eea
where $K_n(x)$ are the modified Bessel functions of the second kind, and the matching results \eqref{mcm} have been used to simplify the expression.

\begin{figure}
\begin{tabular}{cc}
\includegraphics[width=0.49\textwidth]{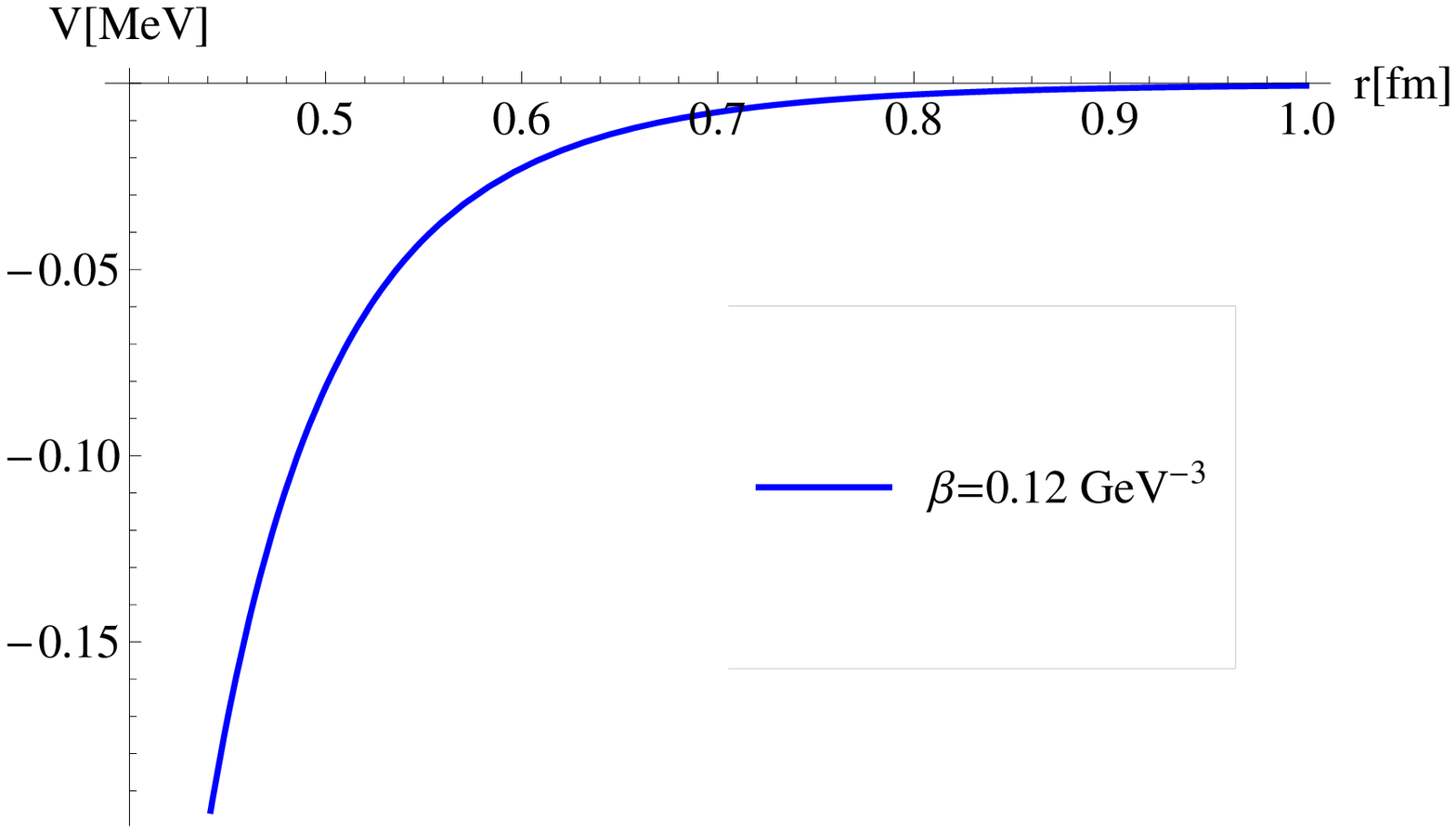} & \includegraphics[width=0.49\textwidth]{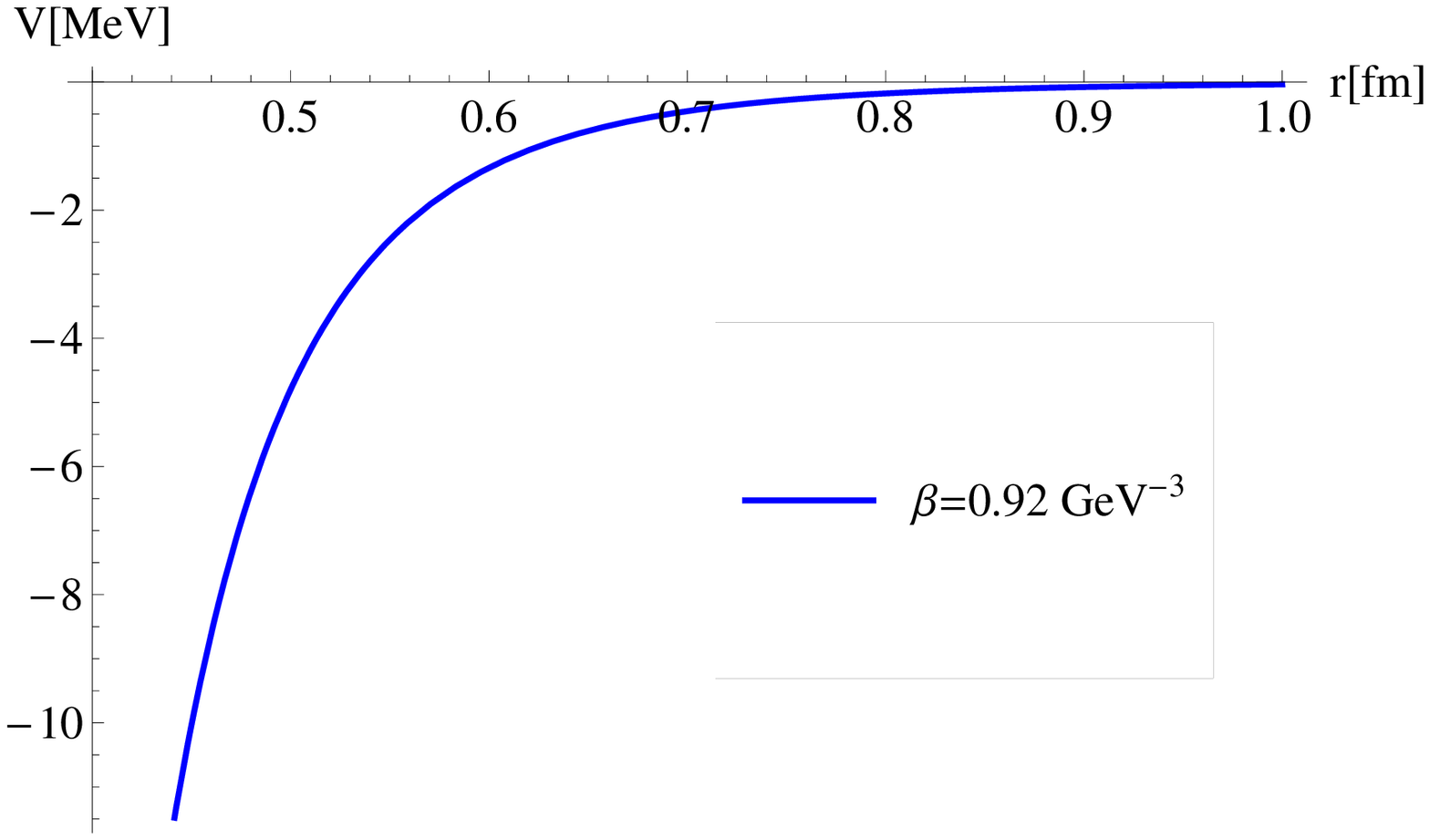} \\
(a) & (b)
\end{tabular}
\caption{Plot of the potential in coordinate space Eq.~\eqref{cspot2} for the smallest and the largest value of the polarizability $\beta$ of $\phi$ quoted in Eq.~\eqref{betaresult}. We take $\kappa=0.186$, $m_{\pi}=135$~MeV and $b$ is computed with three active flavors and $N_c=3$.}
\label{potvsb}
\end{figure}

The potential in coordinate space is plotted in Fig.~\ref{potvsb} for two values of $\beta$. The dependence on $\beta$ is quite noticeable, as one would expect since the potential is proportional to $\beta^2$. For short distances the absolute value of the potential increases rapidly. For the two values of $\kappa$ listed in Sec.~\ref{chiefta} the variation of the potential is unappreciable. Only for very large deviations compared to the uncertainties of these parameters does the potential change in a more significant way and only on the short distances.

\begin{figure}[ht]
\centering{\includegraphics[width=0.49\textwidth]{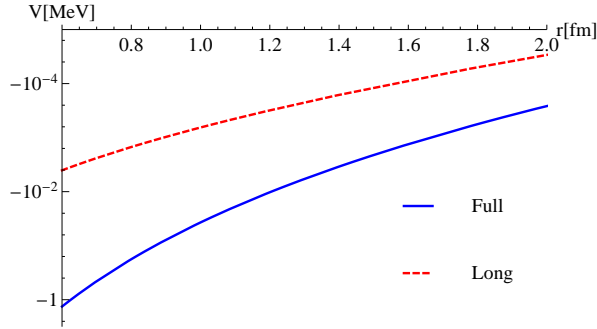}}
\caption{Comparison of the van der Waals potential \eqref{cspot2} (blue line) with its long-range expansion \eqref{longrangevdW} (red line) for $\beta = 0.92$~GeV$^{-3}$ and other parameters like in Fig.~\ref{potvsb}.}
\label{Waalscomp}
\end{figure}

In the long range, i.e., when Eq.~\eqref{cspot2} is expanded for large $r$, we obtain 
\be
W(r)= -\frac{3(3+\kappa_2)^2\pi^{3/2}\beta^2}{4b^2}\frac{m^{9/2}_{\pi}}{r^{5/2}}e^{-2m_{\pi} r}\,.
\label{longrangevdW}
\ee
The long-range potential \eqref{longrangevdW}, valid for $r \gg 1/(2m_\pi)$, is of the order of a few eV, whereas the potential \eqref{cspot2}, valid also for  $r \sim 1/(2m_\pi)$, may be as large as $-1$ MeV in the region around 0.6 and 0.7 fm, see Fig.~\ref{Waalscomp}.

If in the expression \eqref{longrangevdW} we neglect $\kappa$, i.e., we take $\kappa_2=2$, and we also neglect contributions proportional to $m_\pi^2$ in the trace anomaly, then the expression agrees with the one derived in Ref.~\cite{Fujii:1999xn} using an approach similar to ours based on the dipole-dipole interaction and the trace anomaly. One should notice, however, that, while neglecting $\kappa$ is justified by its smallness, taking the chiral limit of the trace anomaly modifies the strength of the long-range van der Waals potential (although not its functional dependence on $r$ and $m_\pi$): the van der Waals potential in Ref.~\cite{Fujii:1999xn} is a factor $16/25$ weaker than Eq.~\eqref{longrangevdW}. This is not surprising if one considers that under the condition that the typical distance between the quarkonia is of order $1/m_\pi$, $m_\pi$ cannot be neglected. In the numerical part of their analysis the authors of Ref.~\cite{Fujii:1999xn} took the polarizability from the large-$N_c$ estimate of Refs.~\cite{Peskin:1979va,Bhanot:1979vb}. 

Comparing the plots of Eq.~\eqref{cspot2} with the analogue results from the two-pion exchange diagrams in the nucleon-nucleon EFT (e.g., in Refs.~\cite{Kaiser:1997mw,Epelbaum:2003gr}) we see that the $\phi$-$\phi$ potential that we have obtained is much less deep. This difference has two origins. First, in the nucleon-nucleon EFT the two-pion exchange appears at next-to-leading order instead of at next-to-next-to-leading order as in the $\phi$-$\phi$ case, this results in a $\mathcal{O}(\left(m_{\pi}/\Lambda_{\chi}\right)^2)\sim 10^{-2}$ suppression. Second, there are five different two-pion exchange diagrams in nucleon-nucleon interactions, whereas in our case there is only one.

\section{Conclusions}
\label{seccon}
The $1S$ states are the lowest lying in the bottomonium spectrum. For these states one can assume that the hierarchy $m v\gg mv^2\gg\Lambda_{\rm QCD}$ is fulfilled. At the ultrasoft energy scale the $1S$ bottomonium states are solutions of the Schr\"odinger equation defined by weakly coupled pNRQCD. Since spin-dependent interactions are suppressed by the bottom quark mass, both $\e$ and $\u$ can be represented by a pseudoscalar field $\phi$. Integrating out the ultrasoft scale we arrive at an EFT in which the color-singlet $1S$ bottomonium states and the gluon fields are dynamical degrees of freedom. In this EFT, which we have named gWEFT, the color-singlet $1S$ bottomonium states interact with the gluons through quadratic terms in the chromoelectric fields proportional to the chromopolarizability of the states. Matching pNRQCD to gWEFT, the chromopolarizability can be computed in perturbation theory. A key ingredient in the calculation of the polarizability is the description of the intermediate color-octet states. In weakly coupled pNRQCD, the octet potential is a Coulomb-like repulsive potential, therefore the octet eigenfunctions correspond to Coulombic wave functions in the continuum region. An expression of the polarizability, where the dependence on the number of colors has been made explicit, is given in Eq.~\eqref{beta2}. The expression agrees with previous findings in the literature.

In gWEFT the gluon dynamics is nonperturbative. To put our results for the chromopolarizability in a more useful form we have used the QCD trace anomaly to obtain the two-pion production amplitude for the quadratic chromoelectric field operator and matched the result to a chiral EFT in which the $1S$ bottomonium and pions are the degrees of freedom. Using this chiral EFT we have computed the leading chiral logarithmic contribution to the $1S$ bottomonium mass. This can be read from Eq.~\eqref{mchirallog}.

The second application we have considered is the calculation of the long-range van der Waals potential generated by the two-pion exchange between two $1S$ bottomonium states. The van der Waals potential is defined at a lower-energy scale than $m_{\pi}$, set by the center-of-mass kinetic energy of the $1S$ bottomonium state. Thus we have written down the EFT at this latter scale, WEFT, and computed the potential as a matching coefficient. Using a dispersive representation of the potential, which takes into account the two-pion cut, an analytical expression of the van der Waals potential has been obtained in Eq.~\eqref{cspot2} for $r \sim 1/(2 m_\pi)$ or larger, which reduces to Eq.~\eqref{longrangevdW} in the limiting case $r \gg 1/(2 m_\pi)$. The results of both applications improve on previous findings.

Our calculation of the $\phi$-$\phi$ long-range potential of Sec.~\ref{lrpot} shows a significant dependence on the value of the polarizability $\beta$. Hence, while the long-range parametric dependence on the distance $r$ of the potential is well understood and resembles that of the two-pion exchange potentials of the nucleon-nucleon EFT, its actual strength reflects the uncertainty on $\beta$. Finally, the possible existence of a shallow $\phi$-$\phi$ bound state will also depend crucially on the $\phi$-$\phi$ short-range interaction, which has not been addressed in the present work.

\bigskip
{\bf Acknowledgments}
\bigskip

N.B. and A.V. acknowledge a suggestion by Estia Eichten to look for a possible $\e$-$\e$ bound state. This work has been supported by the DFG and the NSFC through funds provided to the Sino-German CRC 110 ``Symmetries and the Emergence of Structure in QCD'' and by the DFG cluster of excellence "Origin and Structure of the Universe". A grant from the bilateral agreement between Bayerische Hochschulzentrum f\"ur Lateinamerika (BAYLAT) of the Bayerischen Staatsministeriums für Bildung und Kultus, Wissenschaft und Kunst (StMBW) and Fun\-da\-\c{c}\~ao de Amparo \`a Pesquisa do Estado de S\~ao Paulo (FAPESP), contracts Nos. 914-20.1.3 (BAYLAT) and 2013/50841-1 (FAPESP) is acknowledged. The work of G.K. was partially financed by Conselho Nacional de Desenvolvimento Cient\'{\i}fico e Tecnol\'ogico - CNPq, Grant No. 305894/2009-9, and Fun\-da\-\c{c}\~ao de Amparo \`a Pesquisa do Estado de S\~ao Paulo Grant No. 2013/01907-0.

\appendix

\section{Dispersion relations}
\label{disrel}
\begin{figure}
\centerline{\includegraphics[width=.5\textwidth]{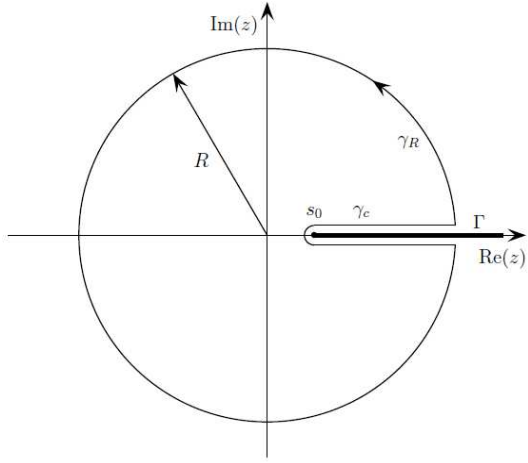}}
\caption{The integration path $\gamma$ consists of a part $\gamma_c$ circumnavigating the branch cut and an arc $\gamma_R$ with radius R.}
\label{cifp}
\end{figure}

Consider a function of one complex variable $f(z)$ that is analytic on the cut complex plane $\mathbb{C}\backslash \Gamma$ with $\Gamma=[s_0,\infty)\subset\mathbb{R}$ and real below the cut: $f(s)\in \mathbb{R}$ $\forall s\in\mathbb{R}$, $s<s_0$. The Schwarz reflection principle holds
\be
f(z^{*})=f^{*}(z) \quad \forall z\in\mathbb{R}\backslash \Gamma\,.
\ee
We can apply Cauchy's integral formula
\be
f(z)=\frac{1}{2\pi i}\oint_{\gamma}\frac{f(\xi)}{\xi-z}\,d\xi
\ee
to the integration path $\gamma$ shown in Fig.~\ref{cifp}.
Assuming that the function $f(z)$ tends to zero for $|z|\rightarrow \infty$, the integral over the arc vanishes for $R\rightarrow \infty$. The integral over $\gamma_c$ remains
\be
\begin{split}
f(z)&=\frac{1}{2\pi i}\int_{\gamma_c}\frac{f(\xi)}{\xi-z}\,d\xi=\lim_{\epsilon\rightarrow0}\frac{1}{2\pi i}\int^{\infty}_{s_0}\frac{f(s+i\epsilon)-f(s-i\epsilon)}{s-z}ds \\
&=\lim_{\epsilon\rightarrow0}\frac{1}{\pi}\int^{\infty}_{s_0}\frac{\text{Im}f(s+i\epsilon)}{s-z}ds\,.
\end{split}
\ee
By evaluating this equation just above the cut, we arrive at
\be
f(s)=\frac{1}{\pi}\int^{\infty}_{s_0}\frac{\text{Im}f(s^{\prime})}{s^{\prime}-s-i\epsilon}ds^{\prime}\,,
\label{dprel}
\ee
where the limit $\epsilon\rightarrow 0$ is understood, and $f(s)$ and $f(s^{\prime})$ are the analytic continuation to the real axis from above the cut. Equations like Eq.~\eqref{dprel} are called dispersion relations. We can calculate the dispersive integral with the help of the identity
\be
\frac{1}{s^{\prime}-s-i\epsilon}=\mathcal{P}\frac{1}{s^{\prime}-s}+i\pi\delta(s^{\prime}-s)\,,
\ee
which means that we transform the dispersion integral into the sum of the Cauchy principal value, $\mathcal{P}$, and $i\pi$ times the residue of the integral.

We are interested in a dispersion relation for potentials in momentum space with a two-pion cut in the negative real axis in the complex three-momentum space. We can obtain such a dispersion relation starting from Eq.~\eqref{dprel} and using $s=-k^2$ and $s^{\prime}=\mu^2$, then we arrive at
\be
f(-k^2)=\frac{2}{\pi}\int^{\infty}_{2m_{\pi}}\frac{\mu\,\text{Im}f(\mu^2+i\epsilon)}{\mu^2+k^2}d\mu\,,
\ee
and by rewriting $f(-k^2)$ as a function of $k$ we arrive at the final form used in Ref.~\cite{Epelbaum:2003gr}
\be
f(k)=\frac{2}{\pi}\int^{\infty}_{2m_{\pi}}\frac{\mu\,\text{Im} f(\epsilon-i\mu)}{\mu^2+k^2}d\mu\,.
\label{esprep}
\ee

In the case where the function $f(z)$ does not fall off fast enough for $z\rightarrow \infty$, or if we simply want to reduce the dependence on $\text{Im}f(s^{\prime})$ at large $s^{\prime}$, we can write a subtracted dispersion relation, i.e., a dispersion relation for the function
\be
g(s)\equiv\frac{f(s)-f(\bar{s})}{s-\bar{s}}\,,
\ee
where $\bar{s}<s_0$ is called the subtraction point. The function $g$ has the same analytical properties as $f$, thus we can write
\be
\frac{f(s)-f(\bar{s})}{s-\bar{s}}=\frac{1}{\pi}\int^{\infty}_{s_0}\frac{1}{s^{\prime}-s-i\epsilon}\text{Im}\left(\frac{f(s^{\prime})-f(\bar{s})}{s^{\prime}-\bar{s}}\right)ds^{\prime}\,,
\label{subt1}
\ee
and since $\text{Im}f(\bar{s})=0$
\be
f(s)=f(\bar{s})+\frac{s-\bar{s}}{\pi}\int^{\infty}_{s_0}\frac{\text{Im}f(s^{\prime})}{(s^{\prime}-\bar{s})(s^{\prime}-s-i\epsilon)}ds^{\prime}\,.
\label{subt11}
\ee
We could repeat this procedure for the function $h(s)\equiv (g(s)-g(\bar{s}_2))/(s-\bar{s}_2)$, $\bar{s}_2<s_0$ to obtain a twice-subtracted dispersion relation and so on. In an $n$-times-subtracted dispersion relation, a polynomial of order $n$ in $s$ multiplies the dispersive integral.

The contribution from the subtraction can be separated from the rest of the dispersive integral by using partial fractioning 
\be
\frac{1}{(s^{\prime}-\bar{s})(s^{\prime}-s-i\epsilon)}=\frac{1}{s-\bar{s}}\left(\frac{1}{s^{\prime}-s-i\epsilon}-\frac{1}{s^{\prime}-\bar{s}}\right)\,,
\ee
in Eq.~\eqref{subt11}
\be
f(s)=f(\bar{s}) - \frac{1}{\pi}\int^{\infty}_{s_0}\frac{\text{Im}f(s^{\prime})}{s^{\prime}-\bar{s}}ds^{\prime}
+ \frac{1}{\pi}\int^{\infty}_{s_0}\frac{\text{Im}f(s^{\prime})}{s^{\prime}-s-i\epsilon}ds^{\prime}\,,
\ee
where the integral of the second term is independent of $s$ and is called a subtraction constant. The subtraction constant can be in general a divergent quantity.

In a physical situation, we can split an amplitude into a polynomial piece and a part that generates the cut in the complex plane. The latter, once appropriately subtracted, can be extended analytically using a dispersion relation. The subtraction constants can then be reabsorbed in suitable redefinitions of the couplings of the theory.


\bibliography{1Sbb-longrange_v4}

\end{document}